\documentclass[aps,preprint,showpacs,groupedaddress]{revtex4-1}
\usepackage{amssymb}
\usepackage{mathrsfs}
\usepackage{amsmath}
\usepackage{mathtools}
\usepackage{pstricks}
\usepackage{color}
\usepackage{graphicx}
\usepackage{amsthm}
\newtheorem*{theorem}{Theorem}
\usepackage{physics}
\begin{document}

%\preprint{}

\title{\bf Exact Lorentz-violating all-loop ultraviolet divergences in scalar field theories}


%\author{William C. Vieira}
%\email{william.vieira@gmail.com}
%\affiliation{\it Departamento de F\'\i sica, Universidade Federal do Piau\'\i, 64049-550, Teresina, PI, Brazil}

\author{P. R. S. Carvalho}
\email{prscarvalho@ufpi.edu.br}
\affiliation{\it Departamento de F\'\i sica, Universidade Federal do Piau\'\i, 64049-550, Teresina, PI, Brazil}

\author{M. I. Sena-Junior}
\email{marconesena@poli.br}
\affiliation{\it Escola Polit\'{e}cnica de Pernambuco, Universidade de Pernambuco, 50720-001, Recife, PE, Brazil}
\affiliation{\it Instituto de F\'{i}sica, Universidade Federal de Alagoas, 57072-900, Macei\'{o}, AL, Brazil}

%\thanks{}


%\date{\today}

\begin{abstract}
In this work we evaluate analytically the ultraviolet divergences of Lorentz-violating massive O($N$) $\lambda\phi^{4}$ scalar field theories, which are exact in the Lorentz-violating mechanism, firstly explicitly at next-to-leading order and latter at any loop level through an induction procedure based on a theorem following from the exact approach, for computing the corresponding critical exponents. For attaining that goal, we employ three different and independent field-theoretic renormalization group methods. The results found for the critical exponents show that they are identical in the three distinct methods and equal to their Lorentz invariant counterparts. Furthermore, we show that the results obtained here, based on the single concept of loop order of the referred terms of the corresponding $\beta$-function and anomalous dimensions, reduce to the ones obtained through the earlier non-exact approach based on a joint redefinition of the field and coupling constant of the theory, in the appropriate limit. 
\end{abstract}

%\pacs{11.30.-j; 64.60.ae; 11.30.Cp}
%\keywords{}

\maketitle


\section{Introduction}\label{Introduction} 

\par  Lorentz symmetry is one of the most fundamental symmetries of nature and the possibility of its violation was theme of great investigation in the last years, usually as a finite perturbative expansion at some Lorentz-violating (LV) parameters and loop number, both in high energy \cite{PhysRevD.92.045016,0004-637X-806-2-269,doi:10.1142/S0217751X15500724,vanTilburg2015236,PhysRevD.86.125015,PhysRevD.84.065030,Carvalho2013850,Carvalho2014320}  as well as in low energy \cite{EurophysLett10821001,doi:10.1142/S0217979215502598,doi:10.1142/S0219887816500493} physics. In the latter realm, the critical exponents were computed, at least, at first order in the Lorentz-violating (LV) parameters $K_{\mu\nu}$ and any loop level for LV scalar field theories \cite{EurophysLett10821001,doi:10.1142/S0217979215502598,doi:10.1142/S0219887816500493}. For that, this evaluation was possible by means of the application of a non-exact approach based on a joint redefinition of the field and coupling constant of the theory. In this work, we present an exact approach, which naturally takes into account the effect of the LV parameters exactly and furthermore for all loop orders. Moreover, we will show that the referred exact approach gives expressions for the $\beta$-function as well as for the corresponding fixed point and anomalous dimensions, besides critical exponents and that these expressions reduce to the ones obtained in the earlier non-exact approach in the appropriate limit.  

\par In this work, we compute analytically the critical exponents for massive O($N$) $\lambda\phi^{4}$ scalar field theories with Lorentz violation. This computation is exact in the LV mechanism. For that, we apply three distinct field-theoretic renormalization group methods and they involve the same theory renormalized at different renormalization schemes. In this field-theoretic formulation, if the critical exponents present the same values when obtained through the three methods, this means that they are universal quantities and we have the confirmation of the universality hypothesis. These universal quantities characterize the critical behavior of distinct systems as a fluid and a ferromagnet. When the critical behavior of two or more distinct systems is characterized by the same critical exponents, we say that they belong to the same universality class. The universality class inspected here is the O($N$) one, which encompasses the particular models: Ising ($N=1$), XY ($N=2$), Heisenberg ($N=3$), self-avoiding random walk ($N=0$) and spherical ($N \rightarrow \infty$) for short-range interactions \cite{Pelissetto2002549}. The critical exponents depend on the dimension $d$ of the system, $N$ and symmetry of some $N$-component order parameter (magnetization for magnetic systems), and if the interactions present are of short- or long-range type. Many works probing the dependence of the critical exponents on the obvious parameters as $d$ \cite{PhysRevB.86.155112,PhysRevE.71.046112} and $N$ \cite{PhysRevLett.110.141601,Butti2005527,PhysRevB.54.7177} were published. Just a few of them were published in the less one, that of symmetry of the order parameter \cite{PhysRevE.78.061124,Trugenberger2005509}. The aim of this work is to probe the exact effect of the LV mechanism on the values for the critical exponents.

\par This paper is organized as follows: In next three Sects., we compute analytically and explicitly the next-to-leading loop order quantum corrections to the critical exponents for LV O($N$) self-interacting $\lambda\phi^{4}$ scalar field taking into account the LV mechanism exactly, by applying three distinct field-theoretic renormalization group methods. In Sect. \ref{Exact Lorentz-violating all-loop order critical exponents} we generalize the results for all loop levels. At the end, we present our conclusions.

\section{Exact Lorentz-violating next-to-leading order critical exponents in the Callan-Symanzik method}\label{Exact Lorentz-violating next-to-leading order critical exponents in the Callan-Symanzik method}

\par We consider a massive LV O($N$) scalar field theory whose bare Lagrangian density in Euclidean spacetime is given by \cite{PhysRevD.84.065030,Carvalho2013850,Carvalho2014320} 
\begin{eqnarray}\label{bare Lagrangian}
\mathscr{L} = \frac{1}{2}(\delta_{\mu\nu} + K_{\mu\nu})\partial^{\mu}\phi_{B}\partial^{\nu}\phi_{B} + \frac{1}{2}m_{B}^{2}\phi_{B}^{2}  +  \frac{\lambda_{B}}{4!}\phi_{B}^{4}.
\end{eqnarray}  
In Eq. above, the bare parameters $\phi_{B}$, $m_{B}$ and $\lambda_{B}$ are the bare field, mass and coupling constant, respectively. The responsible for the symmetry breaking mechanism are the constant symmetric LV coefficients $K_{\mu\nu}$. We can now expand the bare primitively $1$PI vertex parts up to next-to-leading loop order to obtain the desired expansion. But up to this loop order, we have so many diagrams. This number of diagrams can be reduced. We see that the diagrams containing tadpole insertions 
\begin{eqnarray}
\parbox{12mm}{\includegraphics[scale=1.0]{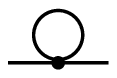}} \parbox{12mm}{\includegraphics[scale=1.0]{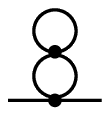}} \parbox{10mm}{\includegraphics[scale=1.0]{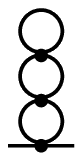}} \parbox{14mm}{\includegraphics[scale=1.0]{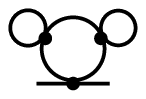}} \parbox{14mm}{\includegraphics[scale=1.0]{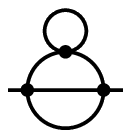}} \parbox{12mm}{\includegraphics[scale=1.0]{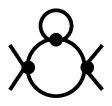}}  \parbox{12mm}{\includegraphics[scale=1.0]{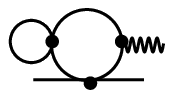}}
\end{eqnarray}
and the one which is independent of external momenta 
\begin{eqnarray}
\parbox{12mm}{\includegraphics[scale=1.0]{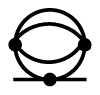}} 
\end{eqnarray}
can be eliminated. It is known that if we substitute the bare mass $m_{B,tree-level}$ in Eq. \ref{bare Lagrangian} initially at tree-level for its three-loop counterpart $m_{B,three-loop}$ \cite{Amit,:/content/aip/journal/jmp/54/9/10.1063/1.4819259} we can achieve the desired aim. Now making $m_{B,three-loop} \rightarrow m_{B}$ from now on we have
\begin{eqnarray}
&&\Gamma_{B}^{(2)}(P^{2} + K_{\mu\nu}P^{\mu}P^{\nu},m_{B},\lambda_{B})  =  \parbox{12mm}{\includegraphics[scale=1.0]{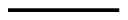}}^{-1}  -  \frac{\lambda_{B}^2}{6}\left( \parbox{12mm}{\includegraphics[scale=1.0]{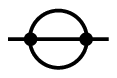}}  - \parbox{12mm}{\includegraphics[scale=1.0]{fig6.eps}}\bigg|_{P^{2} + K_{\mu\nu}P^{\mu}P^{\nu} = 0}\right) + \nonumber \\ &&   \frac{\lambda_{B}^{3}}{4}\left( \parbox{10mm}{\includegraphics[scale=.9]{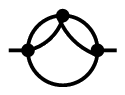}}  - \parbox{10mm} {\includegraphics[scale=.9]{fig7.eps}}\bigg|_{P^{2} + K_{\mu\nu}P^{\mu}P^{\nu} = 0}\right), \label{14}
\end{eqnarray}
\begin{eqnarray}
&&\Gamma_{B}^{(4)}(P_{i},m_{B},\lambda_{B}) =  \lambda_{B} - \frac{\lambda_{B}^{2}}{2}
 \left(\parbox{10mm}{\includegraphics[scale=1.0]{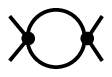}} + 2\hspace{2mm}perm.\right) +   \frac{\lambda_{B}^{3}}{4}
         \left(\parbox{16mm}{\includegraphics[scale=1.0]{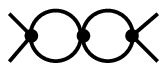}} + 2\hspace{2mm}perm.\right) + \nonumber\\
&& \frac{\lambda_{B}^{3}}{2}
   \left(\parbox{16mm}{\includegraphics[scale=.9]{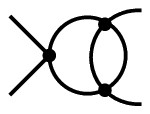}} + 5\hspace{2mm}perm. \right), \label{20}
\end{eqnarray}
\begin{eqnarray}
&& \Gamma_{B}^{(2,1)}(P_{1},P_{2},Q_{3},m_{B},\lambda_{B}) = 1 - 
\frac{\lambda_{B}}{2} \parbox{13mm}{\includegraphics[scale=1.0]{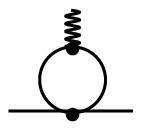}} +  \frac{\lambda_{B}^{2}}{4}\parbox{11mm}{\includegraphics[scale=1.0]{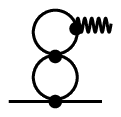}}  + \frac{\lambda_{B}^{2}}{2}\parbox{13mm}{\includegraphics[scale=.9]{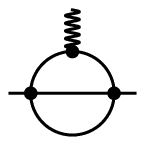}}\label{26}
\end{eqnarray}   
where $Q = -(P_{1} + P_{2})$. We can now define the dimensional and the dimensionless renormalized coupling constants $\lambda$ and $u$ as $\lambda = u m^{\epsilon}$, where $m$, at the loop level considered, is used as an arbitrary momentum scale, thus we can consider the momenta as dimensionless quantities. The same relation between the corresponding bare quantities $\lambda_{B}$ and $u_{0}$ can be also defined as $\lambda_{B} = u_{0}m^{\epsilon}$. We renormalize these correlation functions multiplicatively 
\begin{eqnarray}\label{uhygtfrd}
\Gamma^{(n, l)}(P_{i}, Q_{j}, u, m) = Z_{\phi}^{n/2}Z_{\phi^{2}}^{l}\Gamma_{B}^{(n, l)}(P_{i}, Q_{j}, \lambda_{B}, m_{B})
\end{eqnarray}
which satisfies the Callan-Symanzik equation 
\begin{eqnarray}\label{udhuhsufdhs}
\left( m\frac{\partial}{\partial m} + \beta\frac{\partial}{\partial u} - \frac{1}{2}n\gamma_{\phi} + l\gamma_{\phi^{2}} \right)\Gamma_{R}^{(n, l)}(P_{i}, Q_{j}, u, m) = m^{2}(2 - \gamma_{\phi})\Gamma_{R}^{(n, l + 1)}(P_{i}, Q_{j}, 0, u, m)
\end{eqnarray}
where 
\begin{eqnarray}\label{kjjffxdzs}
\beta(u) = m\frac{\partial u}{\partial m} = -\epsilon\left(\frac{\partial\ln u_{0}}{\partial u}\right)^{-1},
\end{eqnarray}
\begin{eqnarray}\label{koiuhygtf}
\gamma_{\phi}(u) = \beta(u)\frac{\partial\ln Z_{\phi}}{\partial u},
\end{eqnarray}
\begin{eqnarray}\label{koiuhygtfddddd}
\gamma_{\phi^{2}}(u) = -\beta(u)\frac{\partial\ln Z_{\phi^{2}}}{\partial u},
\end{eqnarray}
where we use the function
\begin{eqnarray}\label{udgygeykoiuhygtf}
\overline{\gamma}_{\phi^{2}}(u) = -\beta(u)\frac{\partial\ln \overline{Z}_{\phi^{2}}}{\partial u} \equiv \gamma_{\phi^{2}}(u) - \gamma_{\phi}(u)
\end{eqnarray}
instead of $\gamma_{\phi^{2}}(u)$, for convenience reasons, by fixing the external momenta through the normalization conditions
\begin{eqnarray}\label{ygfdxzsze}
\Gamma^{(2)}(P^{2} + K_{\mu\nu}P^{\mu}P^{\nu} = 0;m,u) = m^{2},
\end{eqnarray}
\begin{eqnarray}
\frac{\partial \Gamma^{(2)}(P^{2} + K_{\mu\nu}P^{\mu}P^{\nu};m^{2},u)}{\partial (P^{2} + K_{\mu\nu}P^{\mu}P^{\nu})}\Biggr|_{P^{2} + K_{\mu\nu}P^{\mu}P^{\nu} = 0} = 1,
\end{eqnarray}
\begin{eqnarray}
\Gamma_{R}^{(4)}(P_{i}=0;m,u) = u,
\end{eqnarray}
\begin{eqnarray}\label{jijhygtfrd}
\Gamma_{R}^{(2,1)}(P_{i}=0,Q_{j}=0,m,u) = 1.
\end{eqnarray}
It is well known that we can reduce more yet the number of diagrams to be evaluated \cite{Amit} because some of them are not independent. This makes this method simpler than the last one which will be applied, the BPHZ one, where, for attaining the same task, we have to compute around of fourteen diagrams. As the computation of the $1$PI vertex parts leads to momentum integration involving just their internal bubbles and not their external legs, all what matters in this evaluation are their internal bubbles contents. Thus, without take into account the O($N$) symmetry factors, we have that \parbox{7mm}{\includegraphics[scale=0.5]{fig10.eps}}$\propto$\hspace{1mm}\parbox{7mm}{\includegraphics[scale=0.5]{fig14.eps}}, \parbox{9mm}{\includegraphics[scale=0.5]{fig21.eps}}$\propto$\hspace{1mm}\parbox{8mm}{\includegraphics[scale=0.5]{fig17.eps}}, \parbox{7mm}{\includegraphics[scale=0.5]{fig11.eps}}\hspace{2mm}$\propto$\hspace{2mm}\parbox{7mm}{\includegraphics[scale=0.5]{fig16.eps}}\hspace{1mm}$\propto$\hspace{2mm}(\parbox{5mm}{\includegraphics[scale=0.5]{fig10.eps}})$^{2}$. Finally, the only diagrams to be evaluated are the $\parbox{6mm}{\includegraphics[scale=0.5]{fig10.eps}}$, $\parbox{8mm}{\includegraphics[scale=0.7]{fig6.eps}}$, $\parbox{7mm}{\includegraphics[scale=0.6]{fig7.eps}}$, $\parbox{7mm}{\includegraphics[scale=0.5]{fig21.eps}}$ ones. Thus we can write the $1$PI vertex parts as
\begin{eqnarray}\label{gtfrdrdes}
\Gamma^{(2)}_{B}(P^{2} + K_{\mu\nu}P^{\mu}P^{\nu}, u_{0}, m_{B}) =  (P^{2} + K_{\mu\nu}P^{\mu}P^{\nu})( 1 - B_{2}u_{0}^{2} + B_{3}u_{0}^{3}),
\end{eqnarray}
\begin{eqnarray}
\Gamma^{(4)}_{B}(P_{i}, u_{0}, m_{B}) =  m_{B}^{\epsilon}u_{0}[ 1 - A_{1}u_{0} + (A_{2}^{(1)} + A_{2}^{(2)})u_{0}^{2}],
\end{eqnarray}
\begin{eqnarray}\label{gtfrdesuuji}
\Gamma^{(2,1)}_{B}(P_{1}, P_{2}, Q_{3}, u_{0}, m_{B}) =  1 - C_{1}u_{0} + (C_{2}^{(1)} + C_{2}^{(2)})u_{0}^{2},
\end{eqnarray}
where
\begin{eqnarray}
A_{1} = \frac{(N + 8)}{6}~\parbox{10mm}{\includegraphics[scale=1.0]{fig10.eps}}_{SP}
\end{eqnarray}
\begin{eqnarray}
A_{2}^{(1)} = \frac{(N^{2} + 6N + 20)}{36}~\parbox{16mm}{\includegraphics[scale=1.0]{fig11.eps}}_{SP},
\end{eqnarray}
\begin{eqnarray}
A_{2}^{(2)} = \frac{(5N + 22)}{9}~\parbox{14mm}{\includegraphics[scale=1.0]{fig21.eps}}_{SP},
\end{eqnarray}
\begin{eqnarray}
B_{2} = \frac{(N + 2)}{18}~\parbox{12mm}{\includegraphics[scale=1.0]{fig6.eps}}^{\prime},
\end{eqnarray}
\begin{eqnarray}
B_{3} = \frac{(N + 2)(N + 8)}{108}~\parbox{12mm}{\includegraphics[scale=1.0]{fig7.eps}}^{\prime},
\end{eqnarray}
\begin{eqnarray}
C_{1} = \frac{(N + 2)}{6}\hspace{1mm}\parbox{14mm}{\includegraphics[scale=1.0]{fig14.eps}}_{SP}, 
\end{eqnarray}
\begin{eqnarray}
C_{2}^{(1)} = \frac{(N + 2)^{2}}{36}\hspace{1mm}\parbox{12mm}{\includegraphics[scale=1.0]{fig16.eps}}_{SP},
\end{eqnarray}
\begin{eqnarray}
C_{2}^{(2)} = \frac{(N + 2)}{6}\hspace{1mm}\parbox{13mm}{\includegraphics[scale=1.0]{fig17.eps}}_{SP}
\end{eqnarray}
and in the Callan-Symanzik method, the two-loop diagram $\parbox{10mm}{\includegraphics[scale=.8]{fig6.eps}}\bigg|_{P^{2} + K_{\mu\nu}P^{\mu}P^{\nu} = 0}$ as well three-loop one given by $\parbox{10mm}{\includegraphics[scale=.7]{fig7.eps}}\bigg|_{P^{2} + K_{\mu\nu}P^{\mu}P^{\nu} = 0}$, do not contribute to the subsequent computations, since we evaluate them at fixed vanishing external momenta. In the right-hand side (rhs) of Eq. \ref{udhuhsufdhs}, the referred $1$PI vertex part is one of $l + 1$ composite field insertions and the one in the left-hand side (lhs) has $l$ such insertions. As it is known, an extra composite field insertion is responsible for one additional power of the propagator in the corresponding $1$PI vertex part. We can then work in the ultraviolet limit, \textit{i. e.}, in the limit where the external momenta $P_{i}/m \rightarrow \infty$. After taking this limit, the rhs can be neglected in comparison with the lhs, order by order in perturbation theory. This is, in essence, the content of the Weinberg's theorem \cite{PhysRev.118.838}. So, the $1$PI vertex parts satisfy the renormalization group equation, thus permitting us to apply the theory of scaling for these functions and evaluate the $\beta$-function and anomalous dimensions as well as the corresponding critical exponents. The LV coefficients can now been considered exactly by noting that $q^{2} + K_{\mu\nu}q^{\mu}q^{\nu} \equiv (\delta_{\mu\nu} + K_{\mu\nu})q^{\mu}q^{\nu}$ = $q^{t}(\mathbb{I}$ + $\mathbb{K})q$, where $q$ is a $d$-dimensional vector whose representation is a column matrix and $q^{t}$ is a row matrix and $\mathbb{I}$ and $\mathbb{K}$ are matrix representations of the identity and $K_{\mu\nu}$, respectively. Thus making $q^{\prime} = \sqrt{\mathbb{I} + \mathbb{K}}\hspace{1mm}q$, the LV mechanism is shown explicitly through two contributions. The first of them is displayed through the volume elements of $d$-dimensional integrals $d^{d}q^{\prime} = \sqrt{det(\mathbb{I} + \mathbb{K})}d^{d}q$, thus $d^{d}q = d^{d}q^{\prime}/\sqrt{det(\mathbb{I} + \mathbb{K})}$. This LV full or exact contribution $\mathbf{\Pi} = 1/\sqrt{det(\mathbb{I} + \mathbb{K})}$ reduces to its perturbative counterpart $\Pi \simeq \Pi^{(0)} + \Pi^{(1)} + \Pi^{(2)}$ for small violations of Lorentz symmetry, where $\Pi^{(i)}$ is the LV contribution of order $i$ in $K_{\mu\nu}$ \cite{PhysRevD.84.065030,Carvalho2013850,Carvalho2014320,EurophysLett10821001,doi:10.1142/S0217979215502598,doi:10.1142/S0219887816500493}. The other LV modification of the theory is that involving the external momenta. It can be seen in the momentum-dependent $d$-dimensional integrals when evaluated in dimensional regularization in $d = 4 - \epsilon$
\begin{eqnarray}
\int \frac{d^{d}q}{(2\pi)^{d}} \frac{1}{(q^{2} + 2Pq + M^{2})^{\alpha}} =  \hat{S}_{d}\frac{1}{2}\frac{\Gamma(d/2)}{\Gamma(\alpha)}\frac{\Gamma(\alpha - d/2)}{(M^{2} - P^{2})^{\alpha - d/2}},   
\end{eqnarray}
where $\hat{S}_{d}=S_{d}/(2\pi)^{d}=2/(4\pi)^{d/2}\Gamma{(d/2)}$, and $S_{d}=2\pi^{d/2}/\Gamma(d/2)$ is the surface area of a unit $d$-dimensional sphere. Its finite value in four-dimensional spacetime is  $\hat{S}_{4}=2/(4\pi)^{2}$. This definition is convenient as to each loop integration we have a factor of $\hat{S}_{4}$ at four dimensions, thus avoiding the appearance of Euler-Mascheroni constants in the middle of calculations \cite{Amit}. Now making $q^{\prime} \rightarrow P^{\prime}$ and $q \rightarrow P$, $P^{\prime 2} = P^{2} + K_{\mu\nu}P^{\mu}P^{\nu}$. As it is known, from all diagrams displayed above, we need to compute only four of them \cite{Amit}. They are shown in \ref{Integrals of Callan-Symanzik method}. When we absorb $\hat{S}$ in a redefinition of the coupling constant and use the Feynman diagrams for computing the $\beta$-function and anomalous dimensions by writing the Laurent expansion 
\begin{eqnarray}\label{suhufjifjvf}
u_{0} = u\left( 1 + \sum_{i=1}^{\infty} a_{i}(\epsilon)u^{i}\right),
\end{eqnarray}
\begin{eqnarray}
Z_{\phi} = 1 + \sum_{i=1}^{\infty} b_{i}(\epsilon)u^{i},
\end{eqnarray}
\begin{eqnarray}\label{iaifkdvkvkck}
\overline{Z}_{\phi^{2}} = 1 + \sum_{i=1}^{\infty} c_{i}(\epsilon)u^{i},
\end{eqnarray}
\begin{eqnarray}
\beta(u) = -\epsilon u[ 1 - a_{1}u + 2(a_{1}^{2} - a_{2})u^{2} ],
\end{eqnarray}
\begin{eqnarray}
\gamma_{\phi}(u) = -\epsilon u[ 2b_{2}u + (3b_{3} - 2b_{2}a_{1})u^{2} ],
\end{eqnarray}
\begin{eqnarray}
\overline{\gamma}_{\phi^{2}}(u) = \epsilon u[ c_{1} + (2c_{2} - c_{1}^{2} - 2a_{1}c_{1})u ],
\end{eqnarray}
where the constant coefficients $a_{1}$, $\cdots$, $c_{2}$ depend on the Feynman diagrams, evaluated in appendix \ref{Integrals of Callan-Symanzik method}, just mentioned \cite{Amit}, we obtain 
\begin{eqnarray}\label{jghyuahuahuahude}
\beta(u) = -\epsilon u + \frac{N + 8}{6}\left( 1 - \frac{1}{2}\epsilon \right)\mathbf{\Pi} u^{2} -  \frac{3N + 14}{12}\mathbf{\Pi}^{2}u^{3},
\end{eqnarray}
\begin{eqnarray}\label{jghyuahuahuahudfrte}
\gamma_{\phi}(u) = \frac{N + 2}{72}\left( 1 - \frac{1}{4}\epsilon + I\epsilon \right)\mathbf{\Pi}^{2}u^{2} - \frac{(N + 2)(N + 8)}{432}(I + 1)\mathbf{\Pi}^{3}u^{3},
\end{eqnarray}
\begin{eqnarray}\label{ujjhhahuahuahuaart}
\overline{\gamma}_{\phi^{2}}(u) = \frac{N + 2}{6}\left( 1 - \frac{1}{2}\epsilon\right)\mathbf{\Pi} u - \frac{N + 2}{12}\mathbf{\Pi}^{2} u^{2}.
\end{eqnarray}
We observe that the expression for the $\beta$-function of Eq. (\ref{jghyuahuahuahude}) can be read off based on a single concept, that of loop order of the referred term of the corresponding function. As we can see, its first term does not originate from a loop integral and the exact approach demands that it has not to be accompanied of a LV full $\mathbf{\Pi}$ factor, although it is a term of first order in $u$. This term is fundamental for making possible expansions in quantum field theory and is essential in the renormalization group and $\epsilon$-expansion techniques developed by Wilson, specially with applications in critical phenomena \cite{Wilson197475,PhysRevLett.28.240,PhysRevLett.28.548} in $d < 4$. Its second one-loop term is of second order in $u$, but it has acquired only a linear power of $\mathbf{\Pi}$. The last one, although being of third order in $u$, must be of second order in $\mathbf{\Pi}$, since it is of two-loop order. Similar arguments can be utilized to the others terms of the anomalous dimensions of Eqs. (\ref{jghyuahuahuahudfrte}) and (\ref{ujjhhahuahuahuaart}) as well. Thus, the exact approach permit us to see that each loop term is accompanied of a power of the LV full $\mathbf{\Pi}$ factor as it is shown by the general theorem displayed in last Sect. This procedure is valid at all intermediate steps of the program. 
%In the earlier non-exact approach, which is based on a joint redefinition of the field $\phi \rightarrow \Pi^{-1/2} \phi$ and coupling constant $u \rightarrow \Pi u$, it is not so evident and direct that the $\beta$-function would result in the expression of Eq. (\ref{jghyuahuahuahude}), in the limit of small $K$. A direct application of the non-exact approach for reading off the $\beta$-function would lead erroneously, for example, to its first term as one of the form $-\epsilon u \rightarrow -\epsilon\Pi u$. Additionally, as its second term is of second order in $u$, it would be erroneously of second order in $\Pi$. The last term would acquire erroneously a cubic power of $\Pi$, since it is of third order in $u$. For obtaining the $\beta$-function of Eq. (\ref{jghyuahuahuahude}), in the limit of small $K$, through the non-exact approach, we must make $u \rightarrow \Pi u$ \emph{just} at the renormalization constants $Z_{\phi}$, $\overline{Z}_{\phi^{2}}$ and $u_{0}$ and no more in the other intermediate steps of that program. Thus, the non-exact approach, applied carefully, have lead, although not so evidently and directly, to the correct earlier LV approximated critical exponents evaluations \cite{EurophysLett10821001,doi:10.1142/S0217979215502598,doi:10.1142/S0219887816500493}. 
Another interesting point to be mentioned is that in this method, the $\beta$-function and anomalous dimensions depend on the LV coefficients at its exact form only through the LV $\mathbf{\Pi}$ factor and on the symmetry point employed. We need to compute the nontrivial solution of the $\beta$-function. The trivial one leads to the mean field or Landau critical exponents and can be obtained mathematically by a factorization procedure resulting in the factorization of a single power of $u$ in Eq. for the $\beta$-function. This procedure results in the nontrivial fixed point given by
\begin{eqnarray}\label{yagyaggyuuhd}
u^{*} = \frac{6\epsilon}{(N + 8)\mathbf{\Pi}}\left\{ 1 + \epsilon\left[ \frac{3(3N + 14)}{(N + 8)^{2}} + \frac{1}{2} \right]\right\}.
\end{eqnarray}
It can be written as $u^{*} = u^{*(0)}/\mathbf{\Pi}$, where $u^{*(0)}$ is its Lorentz-invariant (LI) counterpart. Now the LV corrections to mean field or Landau approximation to the critical exponents are given though the application of definitions $\eta\equiv\gamma_{\phi}(u^{*})$ and $\nu^{-1}\equiv 2 - \eta - \overline{\gamma}_{\phi^{2}}(u^{*})$. They can be applied to obtain, to next-to-leading order, the two respective critical exponents
\begin{eqnarray}\label{eta}
\eta = \frac{(N + 2)\epsilon^{2}}{2(N + 8)^{2}}\left\{ 1 + \epsilon\left[ \frac{6(3N + 14)}{(N + 8)^{2}} -\frac{1}{4} \right]\right\}\equiv\eta^{(0)},
\end{eqnarray}
\begin{eqnarray}\label{nu}
\nu = \frac{1}{2} + \frac{(N + 2)\epsilon}{4(N + 8)} +  \frac{(N + 2)(N^{2} + 23N + 60)\epsilon^{2}}{8(N + 8)^{3}} \equiv\nu^{(0)},
\end{eqnarray}
where $\eta^{(0)}$ and $\nu^{(0)}$ are their corresponding Lorentz-invariant (LI) counterparts \cite{Wilson197475}. As there are six critical exponents and four scaling relations among them, there are only two independent ones. Thus the two ones above are enough for evaluating the four remaining ones. In next Sect. we will attain the same task but now in a distinct renormalization method and will compare the results.

\section{Exact Lorentz-violating next-to-leading order critical exponents in the Unconventional minimal subtraction scheme}\label{Exact Lorentz-violating next-to-leading order critical exponents in the Unconventional minimal subtraction scheme}

\par This method is characterized by its elegance as compared with the earlier one since the external momenta remain at arbitrary values along all the renormalization program. This implies that we do not have to compute any parametric integral because they cancel out in the final expressions for the $\beta$-function and anomalous dimensions. Then, now we have that 
\begin{eqnarray}
A_{1} = \frac{(N + 8)}{18}\left[ \parbox{10mm}{\includegraphics[scale=1.0]{fig10.eps}} + 2 \hspace{2mm} perm. \right]
\end{eqnarray}
\begin{eqnarray}
A_{2}^{(1)} = \frac{(N^{2} + 6N + 20)}{108}\left[ \parbox{10mm}{\includegraphics[scale=1.0]{fig11.eps}}\quad\quad  + 2 \hspace{2mm} perm. \right],
\end{eqnarray}
\begin{eqnarray}
A_{2}^{(2)} = \frac{(5N + 22)}{54}\left[ \parbox{10mm}{\includegraphics[scale=1.0]{fig21.eps}}\quad+ 5 \hspace{2mm} perm. \right],
\end{eqnarray}
\begin{eqnarray}
B_{2} = \frac{(N + 2)}{18}\hspace{1mm}\left(\parbox{12mm}{\includegraphics[scale=1.0]{fig6.eps}} \quad - \quad
% sunset
\parbox{12mm}{\includegraphics[scale=1.0]{fig6.eps}}\bigg|_{P=0}\right),
\end{eqnarray}
\begin{eqnarray}
B_{3} = \frac{(N + 2)(N + 8)}{108}\hspace{1mm}\left(\parbox{12mm}{\includegraphics[scale=1.0]{fig7.eps}} - 
% sunset
\parbox{12mm}{\includegraphics[scale=1.0]{fig7.eps}}\bigg|_{P=0}\right),
\end{eqnarray}
\begin{eqnarray}
C_{1} = \frac{(N + 2)}{6}\hspace{1mm}\parbox{10mm}{\includegraphics[scale=1.0]{fig14.eps}}\hspace{4mm}, 
\end{eqnarray}
\begin{eqnarray}
C_{2}^{(1)} = \frac{(N + 2)^{2}}{36}\hspace{1mm}\parbox{10mm}{\includegraphics[scale=1.0]{fig16.eps}}\hspace{2mm},
\end{eqnarray}
\begin{eqnarray}
C_{2}^{(2)} = \frac{(N + 2)}{6}\hspace{1mm}\parbox{10mm}{\includegraphics[scale=1.0]{fig17.eps}}\hspace{5mm},
\end{eqnarray}
where the poles are minimally eliminated, thus being absorbed in the renormalization constants for the field $Z_{\phi}$ and composite field $Z_{\phi^{2}}$, respectively.  Now absorbing $\hat{S}$ in a redefinition of the coupling constant and using the Feynman diagrams computed in \ref{Integrals of Unconventional minimal subtraction scheme}, we find 
\begin{eqnarray}\label{uahuahuahu}
\beta(u) =  -\epsilon u + \frac{N + 8}{6}\mathbf{\Pi} u^{2} - \frac{3N + 14}{12}\mathbf{\Pi}^{2}u^{3},
\end{eqnarray}
\begin{eqnarray}\label{uahuahuahuwesc}
\gamma_{\phi}(u) = \frac{N + 2}{72}\mathbf{\Pi}^{2}u^{2} - \frac{(N + 2)(N + 8)}{1728}\mathbf{\Pi}^{3}u^{3},
\end{eqnarray}
\begin{eqnarray}\label{uahuahuahuaa}
\overline{\gamma}_{\phi^{2}}(u) = \frac{N + 2}{6}\mathbf{\Pi}u - \frac{N + 2}{12}\mathbf{\Pi}^{2}u^{2}.
\end{eqnarray}
The renormalization program proceeds so elegantly that all the momentum-dependent integrals, namely the ones $L(P^{2} + K_{\mu\nu}P^{\mu}P^{\nu}, m_{B}^{2})$, $L_{3}(P^{2} + K_{\mu\nu}P^{\mu}P^{\nu}, m_{B}^{2})$, $\tilde{i}(P^{2} + K_{\mu\nu}P^{\mu}P^{\nu}, m_{B}^{2})$ have disappeared. Now, the only LV dependence of the theory is that through the LV full $\mathbf{\Pi}$ factor. The cancelling of the integrals aforementioned are associated to the the renormalization of the field and composite field. In fact, technically, the renormalization of these parameters comes from the terms proportional to $P^{2} + K_{\mu\nu}P^{\mu}P^{\nu}$ in the diagrams $\parbox{10mm}{\includegraphics[scale=.8]{fig6.eps}}$ and $\parbox{10mm}{\includegraphics[scale=.7]{fig7.eps}}$. But, unfortunately, we are yet left with a residual divergence and it originates from the terms proportional to $m^{2}$ in the diagrams $\parbox{10mm}{\includegraphics[scale=.8]{fig6.eps}}$ and $\parbox{10mm}{\includegraphics[scale=.7]{fig7.eps}}$. It is show below   
\begin{eqnarray}
&& \Gamma^{(2)}(P^{2} + K_{\mu\nu}P^{\mu}P^{\nu}, u, m) = P^{2} + K_{\mu\nu}P^{\mu}P^{\nu} + \nonumber \\ && m^{2}\left\{1 + \frac{(N+2)}{24}\tilde{I}(P^{2} + K_{\mu\nu}P^{\mu}P^{\nu}, m_{B}^{2}) - \frac{(N+2)(N+8)}{108\epsilon}\tilde{I}(P^{2} + K_{\mu\nu}P^{\mu}P^{\nu}, m_{B}^{2})u^{3} \right\},\quad\quad
\end{eqnarray}
where
\begin{eqnarray}
&&\tilde{I}(P^{2} + K_{\mu\nu}P^{\mu}P^{\nu}, m_{B}^{2}) = \nonumber \\  && \int_{0}^{1} dx \int_{0}^{1}dy lny \frac{d}{dy}\left\{(1-y) \ln\left[\frac{y(1-y)\frac{P^{2} + K_{\mu\nu}P^{\mu}P^{\nu}}{m_{B}^{2}} + 1-y + \frac{y}{x(1-x)}}{1-y + \frac{y}{x(1-x)}} \right]\right\}. 
\end{eqnarray}
The reduction of the number of diagrams to be computed though the redefining of the initial bare mass at tree-level to its three-loop order counterpart produces this residual divergence. We can overcome this problem then subtracting this pole minimally by redefining the two-point function as   
\begin{eqnarray}
&& \tilde{\Gamma}^{(2)}(P^{2} + K_{\mu\nu}P^{\mu}P^{\nu}, u, m) = \nonumber \\  && \Gamma^{(2)}(P^{2} + K_{\mu\nu}P^{\mu}P^{\nu}, u, m) +  m^{2}\left\{ \frac{(N+2)(N+8)}{108\epsilon}\tilde{I}(P^{2} + K_{\mu\nu}P^{\mu}P^{\nu}, m_{B}^{2}) u^{3} \right\}.
\end{eqnarray}
This turn out to connect the Unconventional minimal subtraction scheme to the conventional one in the massless theory \cite{Amit}, once the terms proportional to $m^{2}$ vanish in the latter case. The final check of this redefinition can be shown by showing that it satisfies the normalization condition used in Sect. \ref{Exact Lorentz-violating next-to-leading order critical exponents in the Callan-Symanzik method}
\begin{eqnarray}
&& \tilde{\Gamma}^{(2)}(P^{2} + K_{\mu\nu}P^{\mu}P^{\nu} = 0, u, m) = \nonumber \\ && \Gamma^{(2)}(P^{2} + K_{\mu\nu}P^{\mu}P^{\nu} = 0, u, m) = m^{2}.
\end{eqnarray}
Once again, for computing the LV loop quantum corrections to the critical exponents, we need to evaluate the nontrivial fixed point though the nontrivial solution for the equation $\beta(u^{*}) = 0$. It is given by   
\begin{eqnarray}\label{yagyaguhd}
u^{*} = \frac{6\epsilon}{(N + 8)\Pi}\left\{ 1 + \epsilon\left[ \frac{3(3N + 14)}{(N + 8)^{2}} \right]\right\}.
\end{eqnarray}
This value for the nontrivial fixed point when used for evaluating the critical exponents, leads to the same ones of the earlier Sect.. On more time we confirm the universality hypothesis, that the critical exponents are universal quantities, thus being the same when obtained in different renormalization schemes. Now we proceed to compute the critical exponents in a third renormalization scheme.

\section{Exact Lorentz-violating next-to-leading order critical exponents in the BPHZ method}\label{Exact Lorentz-violating next-to-leading order critical exponents in the BPHZ method}

\par The BPHZ (Bogoliubov-Parasyuk-Hepp-Zimmermann) method \cite{BogoliubovParasyuk,Hepp,Zimmermann} is the most general from all known renormalization methods. It does not include any trick for reducing the total number of diagrams to be evaluated. Thus we have to compute all diagrams in the original expansion for a given loop order. As opposed to the earlier ones, in the BPHZ method, we start from the renormalized theory
\begin{eqnarray}\label{huytrjii}
\mathcal{L} = \frac{1}{2}Z_{\phi}(g_{\mu\nu} + K_{\mu\nu})\partial^{\mu}\phi\partial^{\nu}\phi + \frac{\mu^{\epsilon}u}{4!}Z_{u}\phi^{4} + \frac{1}{2}tZ_{\phi^{2}}\phi^{2},\nonumber \\
\end{eqnarray}
where
\begin{eqnarray}\label{huytr}
\phi = Z_{\phi}^{-1/2}\phi_{B}, \hspace{5mm} u = \mu^{-\epsilon}\frac{Z_{\phi}^{2}}{Z_{u}}\lambda_{B},  \hspace{5mm} t = \frac{Z_{\phi}}{Z_{\phi^{2}}}t_{B}. 
\end{eqnarray}
Initially, considering the bare theory at one-loop order, we absorb that divergence by adding terms to the initial Lagrangian density. Then, a finite Lagrangian density is found. For considering the bare theory at the next loop level, we apply the same procedure and so on, order by order in perturbation theory. Thus we absorb the divergences in the renormalization constants. We expand the renormalization constants as
\begin{eqnarray}\label{uhguhfgugu}
Z_{\phi} = 1 + \sum_{i=1}^{\infty} c_{\phi}^{i},  
\end{eqnarray} 
\begin{eqnarray}
Z_{u} = 1 + \sum_{i=1}^{\infty} c_{u}^{i},
\end{eqnarray}
\begin{eqnarray}
Z_{m^{2}} = 1 + \sum_{i=1}^{\infty} c_{m^{2}}^{i}.
\end{eqnarray}
The $c_{\phi}^{i}$, $c_{g}^{i}$ and $c_{m}^{i}$ coefficients are the $i$-th loop order renormalization constants for the field, renormalized coupling constant and composite field, respectively. They are given by
\begin{eqnarray}\label{Zphi}
Z_{\phi}(u,\epsilon^{-1}) = 1 + \frac{1}{P^{2} + K_{\mu\nu}P^{\mu}P^{\nu}} \Biggl[ \frac{1}{6} \mathcal{K} 
\left(\parbox{12mm}{\includegraphics[scale=1.0]{fig6.eps}}
\right) \Biggr|_{m^2 = 0} S_{\parbox{10mm}{\includegraphics[scale=0.5]{fig6.eps}}} + \nonumber \\   \frac{1}{4} \mathcal{K} 
\left(\parbox{12mm}{\includegraphics[scale=1.0]{fig7.eps}} \right) \Biggr|_{m^2 = 0} S_{\parbox{6mm}{\includegraphics[scale=0.5]{fig7.eps}}} + \frac{1}{3} \mathcal{K}
  \left(\parbox{12mm}{\includegraphics[scale=1.0]{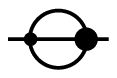}} \right) S_{\parbox{6mm}{\includegraphics[scale=0.5]{fig26.eps}}} \Biggr], \hspace{4mm}
\end{eqnarray}

\begin{eqnarray}\label{Zg}
&&Z_{u}(u,\epsilon^{-1}) = 1 + \frac{1}{\mu^{\epsilon}u} \Biggl[ \frac{1}{2} \mathcal{K} 
\left(\parbox{10mm}{\includegraphics[scale=1.0]{fig10.eps}} + 2 \hspace{1mm} perm.
\right) S_{\parbox{10mm}{\includegraphics[scale=0.5]{fig10.eps}}} +  \frac{1}{4} \mathcal{K} 
\left(\parbox{17mm}{\includegraphics[scale=1.0]{fig11.eps}} + 2 \hspace{1mm} perm. \right) S_{\parbox{10mm}{\includegraphics[scale=0.5]{fig11.eps}}} + \nonumber \\ &&  \frac{1}{2} \mathcal{K} 
\left(\parbox{12mm}{\includegraphics[scale=.8]{fig21.eps}} + 5 \hspace{1mm} perm. \right) S_{\parbox{10mm}{\includegraphics[scale=0.4]{fig21.eps}}} +  \frac{1}{2} \mathcal{K} 
\left(\parbox{10mm}{\includegraphics[scale=1.0]{fig13.eps}} + 2 \hspace{1mm} perm.
\right) S_{\parbox{10mm}{\includegraphics[scale=0.5]{fig13.eps}}} + \nonumber \\ && \mathcal{K}
  \left(\parbox{10mm}{\includegraphics[scale=1.0]{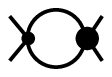}} + 2 \hspace{1mm} perm. \right) S_{\parbox{6mm}{\includegraphics[scale=0.5]{fig25.eps}}} +  \mathcal{K}
  \left(\parbox{10mm}{\includegraphics[scale=1.0]{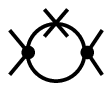}} + 2 \hspace{1mm} perm. \right) S_{\parbox{6mm}{\includegraphics[scale=0.5]{fig24.eps}}} \Biggr],
\end{eqnarray}

\begin{eqnarray}\label{Zphi}
&& Z_{m^{2}}(u,\epsilon^{-1}) = 1 + \frac{1}{m^{2}} \Biggl[ \frac{1}{2} \mathcal{K} 
\left(\parbox{12mm}{\includegraphics[scale=1.0]{fig1.eps}}
\right) S_{\parbox{10mm}{\includegraphics[scale=0.5]{fig1.eps}}} + \frac{1}{4} \mathcal{K} 
\left(\parbox{12mm}{\includegraphics[scale=1.0]{fig2.eps}} \right) S_{\parbox{6mm}{\includegraphics[scale=0.5]{fig2.eps}}} + \frac{1}{2} \mathcal{K}
  \left(\parbox{12mm}{\includegraphics[scale=1.0]{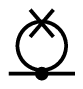}} \right) S_{\parbox{6mm}{\includegraphics[scale=0.5]{fig22.eps}}} + \nonumber \\ && \frac{1}{2} \mathcal{K}
  \left(\parbox{12mm}{\includegraphics[scale=1.0]{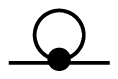}} \right) S_{\parbox{6mm}{\includegraphics[scale=0.5]{fig23.eps}}} + \frac{1}{6} \mathcal{K}
  \left(\parbox{12mm}{\includegraphics[scale=1.0]{fig6.eps}} \right)\Biggr|_{P^2 + K_{\mu\nu}P^{\mu}P^{\nu} = 0} S_{\parbox{6mm}{\includegraphics[scale=0.5]{fig6.eps}}} \Biggr], \hspace{4mm}
\end{eqnarray}
where $S_{\parbox{6mm}{\includegraphics[scale=0.5]{fig6.eps}}}$ is the symmetry factor for the corresponding diagram and so on when some $N$-component field is considered. By using the diagrams in \ref{Integrals of BPHZ method}, we have that the $\beta$-function and anomalous dimensions are given by
\begin{eqnarray}\label{reewriretjgjk}
\beta(u) = -\epsilon u + \frac{N + 8}{6}\mathbf{\Pi} u^{2} - \frac{3N + 14}{12}\mathbf{\Pi}^{2}u^{3},
\end{eqnarray} 
\begin{eqnarray}\label{jkjkpfgjrftj}
\gamma_{\phi}(u) = \frac{N + 2}{72}\mathbf{\Pi}^{2}u^{2} - \frac{(N + 2)(N + 8)}{1728}\mathbf{\Pi}^{3}u^{3},
\end{eqnarray} 
\begin{eqnarray}\label{gfydsguyfsdgufa}
\gamma_{m^{2}}(u) = \frac{N + 2}{6}\mathbf{\Pi} u - \frac{5(N + 2)}{72}\mathbf{\Pi}^{2}u^{2}.
\end{eqnarray} 
One more time, we have to compute the nontrivial solution of Eq. (\ref{reewriretjgjk}). This procedure yields the value 
\begin{eqnarray}\label{fghyagyaguhd}
u^{*} = \frac{6\epsilon}{(N + 8)\mathbf{\Pi}}\left\{ 1 + \epsilon\left[ \frac{3(3N + 14)}{(N + 8)^{2}} \right]\right\}.
\end{eqnarray}
Now by applying the relations $\eta\equiv \gamma_{\phi}(u^{*})$ and $\nu^{-1}\equiv 2 - \gamma_{m^{2}}(u^{*})$, we obtain once again that the LV critical exponents are identical to their LV counterparts. Now we evaluate the LV critical exponents for any loop levels.

\section{Exact Lorentz-violating all-loop order critical exponents}\label{Exact Lorentz-violating all-loop order critical exponents}

\par For computing the critical exponents for all loop levels, we can employ any of the methods aforementioned since the critical exponents, as being universal quantities, must be the same if evaluated at any renormalization scheme. For that, we will employ the BPHZ method which is the most general one. Before that, we need to assert the following theorem 
\begin{theorem} 
Consider a given Feynman diagram in momentum space of any loop order in a theory represented by the Lagrangian density of Eq. (\ref{bare Lagrangian}). Its evaluated expression in dimensional regularization in $d = 4 - \epsilon$ can be written as a general functional $\mathbf{\Pi}^{L}\mathcal{F}(u,P^{2} + K_{\mu\nu}P^{\mu}P^{\nu},\epsilon,\mu)$ if its LI counterpart is given by $\mathcal{F}(u,P^{2},\epsilon,\mu,m)$, where $L$ is the number of loops of the corresponding diagram.
\end{theorem}

\begin{proof} 
A general Feynman diagram of loop level $L$ is a multidimensional integral in $L$ distinct and independent momentum integration variables $q_{1}$, $q_{2}$,...,$q_{L}$, each one with volume element given by $d^{d}q_{i}$ ($i = 1, 2,...,L$). As showed in last Section, the substitution $q^{\prime} = \sqrt{\mathbb{I} + \mathbb{K}}\hspace{1mm}q$ transforms each volume element as $d^{d}q^{\prime} = \sqrt{det(\mathbb{I} + \mathbb{K})}d^{d}q$. Thus $d^{d}q = d^{d}q^{\prime}/\sqrt{det(\mathbb{I} + \mathbb{K})} \equiv \mathbf{\Pi}d^{d}q^{\prime}$, $\mathbf{\Pi} = 1/\sqrt{det(\mathbb{I} + \mathbb{K})}$. Then, the integration in $L$ variables results in a LV overall factor of $\mathbf{\Pi}^{L}$. Now making $q^{\prime} \rightarrow P^{\prime}$ in the substitution above, where $P^{\prime}$ is the transformed external momentum, then $P^{\prime 2} = P^{2} + K_{\mu\nu}P^{\mu}P^{\nu}$. So a given Feynman diagram, evaluated in dimensional regularization in $d = 4 - \epsilon$, assumes the expression $\mathbf{\Pi}^{L}\mathcal{F}(u,P^{2} + K_{\mu\nu}P^{\mu}P^{\nu},\epsilon,\mu)$, where $\mathcal{F}$ is associated to the corresponding diagram if the LI Feynman diagram counterpart evaluation results in $\mathcal{F}(u,P^{2},\epsilon,\mu)$. 
\end{proof} 
Now using the result of the theorem above and the one in which all momentum-dependent integrals cancel out order by order in perturbation theory for all levels in the renormalization process\cite{BogoliubovParasyuk,Hepp,Zimmermann}, we see that the only LV dependence of $\beta$-function and anomalous dimensions is due to the LV full $\mathbf{\Pi}$ factor, which comes from the volume elements of the diagrams contributing with a $\mathbf{\Pi}^{L}$ factor, where $L$ is the number of loops of the corresponding graph. Thus we can write the $\beta$-function and anomalous dimensions for all loop levels
\begin{eqnarray}\label{uhgufhduhufdhu}
\beta(u) =  -\epsilon u + \sum_{n=2}^{\infty}\beta_{n}^{(0)}\mathbf{\Pi}^{n-1}u^{n}, 
\end{eqnarray}
\begin{eqnarray}
\gamma(u) = \sum_{n=2}^{\infty}\gamma_{n}^{(0)}\mathbf{\Pi}^{n}u^{n},
\end{eqnarray}
\begin{eqnarray}
\gamma_{m^{2}}(u) = \sum_{n=1}^{\infty}\gamma_{m^{2}, n}^{(0)}\mathbf{\Pi}^{n}u^{n},
\end{eqnarray}
where $\beta_{n}^{(0)}$, $\gamma_{n}^{(0)}$ and $\gamma_{m^{2}, n}$ are the LI nth-loop corrections to the referred functions. By applying the same factorization process employed in the finite loop scenario for the any loop realm, we obtain that $u^{*} = u^{*(0)}/\mathbf{\Pi}$ where $u^{*(0)}$ is the LI fixed point for all loop levels. Then, we can substitute this all-loop order fixed point in the $\beta$-function and anomalous dimensions to obtain the LV critical exponents valid for any loop levels as being identical to their any loop orders LI counterparts.

\section{Conclusions}\label{Conclusions}

\par We have evaluated analytically the ultraviolet divergences of Lorentz-violating massive O($N$) $\lambda\phi^{4}$ scalar field theories, which are exact in the Lorentz-violating mechanism, firstly explicitly at next-to-leading order and latter at any loop level through an induction procedure based on a theorem following from the exact approach, for computing the corresponding critical exponents. For that, we have employed three different and independent field-theoretic renormalization group methods. We have found equal critical exponents in the three methods and furthermore identical to their Lorentz invariant counterparts. We have also showed that the exact approach, which reduces to the non-exact one in its limited range of applicability, besides exact, is capable of furnishing the expressions for the all-loop LV radiative quantum corrections to the $\beta$-function and anomalous dimensions considering just a single concept, that of loop number of the corresponding terms of these functions. Furthermore, the present exact approach, when applied to the referred theory, is the first one in literature for our knowledge. Thus it can inspire the exact solution of problems involving considering the exact effect of LV mechanisms in many physical phenomena ranging from high- (standard model extension for example) to low energy physics (corrections to scaling, finite-size scaling, amplitude ratios, critical exponents in geometries subjected to different boundary conditions, Lifshitz points etc \cite{Phys.Rev.B.67.104415,Phys.Rev.B.72.224432,Carvalho2009178,Carvalho2010151}.

\appendix

\section{Integrals of Callan-Symanzik method}\label{Integrals of Callan-Symanzik method}

\begin{eqnarray}
&&\parbox{10mm}{\includegraphics[scale=1.0]{fig10.eps}}_{SP} = \frac{1}{\epsilon}\left(1 - \frac{1}{2}\epsilon \right)\mathbf{\Pi},
\end{eqnarray}   
\begin{eqnarray}
&&\parbox{12mm}{\includegraphics[scale=1.0]{fig6.eps}}^{\prime} = -\frac{1}{8\epsilon}\left( 1 - \frac{1}{4}\epsilon +I\epsilon \right)\mathbf{\Pi}^{2},
\end{eqnarray}  
\begin{eqnarray}
&&\parbox{12mm}{\includegraphics[scale=0.9]{fig7.eps}}^{\prime} = -\frac{1}{6\epsilon^{2}}\left( 1 - \frac{1}{4}\epsilon +\frac{3}{2}I\epsilon \right)\mathbf{\Pi}^{3},
\end{eqnarray}  
\begin{eqnarray}
&&\parbox{12mm}{\includegraphics[scale=0.8]{fig21.eps}}_{SP} = \frac{1}{2\epsilon^{2}}\left(1 - \frac{1}{2}\epsilon \right)\mathbf{\Pi}^{2},
\end{eqnarray}  
where the integral $I$ \cite{PhysRevD.8.434,Carvalho2009178,Carvalho2010151}
\begin{eqnarray}
&& I = \int_{0}^{1} \left\{ \frac{1}{1 - x(1 - x)} + \frac{x(1 - x)}{[1 - x(1 - x)]^{2}}\right\}
\end{eqnarray}
is a residual number and is a consequence of the symmetry point chosen.

\section{Integrals of Unconventional minimal subtraction scheme}\label{Integrals of Unconventional minimal subtraction scheme}

\begin{eqnarray}
\parbox{10mm}{\includegraphics[scale=1.0]{fig10.eps}} = \frac{1}{\epsilon} \left[1 - \frac{1}{2}\epsilon -  \frac{1}{2}\epsilon L(P^{2} + K_{\mu\nu}P^{\mu}P^{\nu}, m_{B}^{2}) \right]\mathbf{\Pi},
\end{eqnarray}   

\begin{eqnarray}
&&\parbox{12mm}{\includegraphics[scale=1.0]{fig6.eps}} = \left\{-\frac{3 m_{B}^{2}}{2 \epsilon^{2}}\left[1 + \frac{1}{2}\epsilon + \left(\frac{\pi^{2}}{12} +1 \right)\epsilon^{2} \right] -  \frac{3 m_{B}^{2}}{4}\tilde{i}(P^{2} + K_{\mu\nu}P^{\mu}P^{\nu}, m_{B}^{2})  - \right.  \nonumber \\  &&\left. \frac{(P^{2} + K_{\mu\nu}P^{\mu}P^{\nu})}{8 \epsilon} \left[1 + \frac{1}{4}\epsilon - 2 \epsilon L_{3}(P^{2} + K_{\mu\nu}P^{\mu}P^{\nu},m_{B}^{2})\right]\right\}\mathbf{\Pi}^{2}, 
\end{eqnarray}
\begin{eqnarray}
&&\parbox{12mm}{\includegraphics[scale=1.0]{fig7.eps}} = \left\{-\frac{5 m_{B}^{2}}{3 \epsilon^{3}}\left[1 + \epsilon + \left(\frac{\pi^{2}}{24} + \frac{15}{4} \right)\epsilon^{2} \right] - \frac{5 m_{B}^{2}}{2 \epsilon}\tilde{i}(P^{2} + K_{\mu\nu}P^{\mu}P^{\nu}, m_{B}^{2}) \right.  \nonumber \\  &&\left. -\frac{P^{2} + K_{\mu\nu}P^{\mu}P^{\nu}}{6 \epsilon^{2}} \left[1+ \frac{1}{2}\epsilon - 3 \epsilon L_{3}(P^{2} + K_{\mu\nu}P^{\mu}P^{\nu},m_{B}^{2})\right]\right\}\mathbf{\Pi}^{3}, 
\end{eqnarray}
\begin{eqnarray}
\parbox{14mm}{\includegraphics[scale=1.0]{fig21.eps}} = \frac{1}{\epsilon^{2}} \left[1 - \frac{1}{2}\epsilon - \epsilon L(P^{2} + K_{\mu\nu}P^{\mu}P^{\nu}, m_{B}^{2}) \right]\mathbf{\Pi}^{2},
\end{eqnarray}  
where
\begin{eqnarray}
L(P^{2} + K_{\mu\nu}P^{\mu}P^{\nu}, m_{B}^{2}) = \int_{0}^{1}dx\ln[x(1-x)(P^{2} + K_{\mu\nu}P^{\mu}P^{\nu}) + m_{B}^{2}],
\end{eqnarray}
\begin{eqnarray}
L_{3}(P^{2} + K_{\mu\nu}P^{\mu}P^{\nu}, m_{B}^{2}) = \int_{0}^{1}dx(1-x)  \ln[x(1-x)(P^{2} + K_{\mu\nu}P^{\mu}P^{\nu}) + m_{B}^{2}],
\end{eqnarray}
\begin{eqnarray}
&&\tilde{i}(P^{2} + K_{\mu\nu}P^{\mu}P^{\nu}, m_{B}^{2}) = \nonumber \\ && \int_{0}^{1} dx \int_{0}^{1}dy  \ln y \frac{d}{dy}\left((1-y) \ln\left\{y(1-y)P^{2} + \left[1-y + \frac{y}{x(1-x)}\right]m_{B}^{2} \right\}\right),
\end{eqnarray}

\section{Integrals of BPHZ method}\label{Integrals of BPHZ method}

\begin{eqnarray}
\left(\parbox{12mm}{\includegraphics[scale=1.0]{fig6.eps}}\right)\Biggr|_{m^{2}=0} =  -\frac{u^{2}(P^2 + K_{\mu\nu}P^{\mu}P^{\nu})}{8\epsilon} \left[ 1 + \frac{1}{4}\epsilon -2\epsilon\, J_{3}(P^{2} + K_{\mu\nu}P^{\mu}P^{\nu}) \right]\mathbf{\Pi}^{2},
\end{eqnarray}
\begin{eqnarray}
\parbox{12mm}{\includegraphics[scale=1.0]{fig7.eps}}\bigg|_{m^{2}=0} = \frac{(P^2 + K_{\mu\nu}P^{\mu}P^{\nu})u^{3}}{6\epsilon^{2}} \left[1 + \frac{1}{2}\epsilon - 3\epsilon\, J_{3}(P^{2} + K_{\mu\nu}P^{\mu}P^{\nu})\right]\mathbf{\Pi}^{3},
\end{eqnarray}
\begin{eqnarray}
\parbox{10mm}{\includegraphics[scale=1.0]{fig26.eps}} \quad = -\frac{3(P^{2} + K_{\mu\nu}P^{\mu}P^{\nu})u^{3}}{16\epsilon^{2}}\left[1 + \frac{1}{4}\epsilon - 2\epsilon\, J_{3}(P^{2} + K_{\mu\nu}P^{\mu}P^{\nu})\right]\mathbf{\Pi}^{3},
\end{eqnarray}
\begin{eqnarray}
\parbox{10mm}{\includegraphics[scale=1.0]{fig10.eps}} = \frac{\mu^{\epsilon}u^{2}}{\epsilon} \left[1 - \frac{1}{2}\epsilon - \frac{1}{2}\epsilon J(P^{2} + K_{\mu\nu}P^{\mu}P^{\nu}) \right]\mathbf{\Pi},
\end{eqnarray}
\begin{eqnarray}
\parbox{16mm}{\includegraphics[scale=1.0]{fig11.eps}} = -\frac{\mu^{\epsilon}u^{3}}{\epsilon^{2}} \left[1 - \epsilon - \epsilon J(P^{2} + K_{\mu\nu}P^{\mu}P^{\nu}) \right]\mathbf{\Pi}^{2},\quad\quad
\end{eqnarray}
\begin{eqnarray}
\parbox{12mm}{\includegraphics[scale=0.8]{fig21.eps}} = -\frac{2\mu^{\epsilon}u^{3}}{2\epsilon^{2}} \left[1 - \frac{1}{2}\epsilon - \epsilon J(P^{2} + K_{\mu\nu}P^{\mu}P^{\nu}) \right]\mathbf{\Pi}^2, \quad\quad
\end{eqnarray}
\begin{eqnarray}
\parbox{12mm}{\includegraphics[scale=1.0]{fig13.eps}} =   \frac{\mu^{\epsilon}u^{3}}{2\epsilon^{2}} J_{4}(P^{2} + K_{\mu\nu}P^{\mu}P^{\nu})\mathbf{\Pi}^{2},
\end{eqnarray}
\begin{eqnarray}
\parbox{10mm}{\includegraphics[scale=1.0]{fig25.eps}} = \frac{3\mu^{\epsilon}u^{3}}{2\epsilon^{2}} \left[1 - \frac{1}{2}\epsilon - \frac{1}{2}\epsilon J(P^{2} + K_{\mu\nu}P^{\mu}P^{\nu}) \right]\mathbf{\Pi}^{2},\quad\quad
\end{eqnarray}
\begin{eqnarray}
\parbox{12mm}{\includegraphics[scale=1.0]{fig24.eps}} =  -\frac{\mu^{\epsilon}u^{3}}{2\epsilon^{2}} J_{4}(P^{2} + K_{\mu\nu}P^{\mu}P^{\nu})\,\mathbf{\Pi}^{2},
\end{eqnarray}
\begin{eqnarray}
\parbox{12mm}{\includegraphics[scale=1.0]{fig1.eps}} =
\frac{m^{2}u}{(4\pi)^{2}\epsilon}\left[ 1 - \frac{1}{2}\epsilon\ln\left(\frac{m^{2}}{4\pi\mu^{2}}\right)\right]\mathbf{\Pi},
\end{eqnarray}
\begin{eqnarray}
\parbox{8mm}{\includegraphics[scale=1.0]{fig2.eps}} = - \frac{m^{2}u^{2}}{(4\pi)^{4}\epsilon^{2}}\left[ 1 - \frac{1}{2}\epsilon - \epsilon\ln\left(\frac{m^{2}}{4\pi\mu^{2}}\right)\right]\,\mathbf{\Pi}^{2},
\end{eqnarray}
\begin{eqnarray}
\parbox{12mm}{\includegraphics[scale=1.0]{fig22.eps}} =  \frac{m^{2}g^{2}}{2\epsilon^{2}}\left[ 1 - \frac{1}{2}\epsilon - \frac{1}{2} \epsilon\ln\left(\frac{m^{2}}{4\pi\mu^{2}}\right)\right]\mathbf{\Pi}^{2},
\end{eqnarray}
\begin{eqnarray}
\parbox{12mm}{\includegraphics[scale=1.0]{fig23.eps}} =  \frac{3m^{2}u^{2}}{2\epsilon^{2}}\left[ 1 - \frac{1}{2} \epsilon\ln\left(\frac{m^{2}}{4\pi\mu^{2}}\right)\right]\mathbf{\Pi}^{2},
\end{eqnarray}
\begin{eqnarray}
\left(\parbox{12mm}{\includegraphics[scale=1.0]{fig6.eps}}\right)\Biggr|_{P^{2} + K_{\mu\nu}P^{\mu}P^{\nu}=0} =   -\frac{3m^{2}g^{2}}{2\epsilon}\left[ 1 + \frac{1}{2}\epsilon -\epsilon\ln\left(\frac{m^{2}}{4\pi\mu^{2}}\right)\right]\mathbf{\Pi}^{2},
\end{eqnarray}
where
\begin{eqnarray}\label{uhduhufgjg}
J(P^{2} + K_{\mu\nu}P^{\mu}P^{\nu}) = \int_{0}^{1} dx \ln \left[\frac{x(1-x)(P^{2}+ K_{\mu\nu}P^{\mu}P^{\nu}) + m^{2}}{\mu^{2}}\right],\quad\quad
\end{eqnarray}
\begin{eqnarray}\label{uhduhufgjgdhg}
&& J_{3}(P^{2} + K_{\mu\nu}P^{\mu}P^{\nu}) = \int_{0}^{1}\int_{0}^{1} dx \, dy\,(1-y)\times \nonumber \\ && \ln \Biggl\{\frac{y(1-y)(P^{2}+ K_{\mu\nu}P^{\mu}P^{\nu})}{\mu^{2}} + \left[1-y + \frac{y}{x(1-x)}  \right]\frac{m^{2}}{\mu^{2}}\Biggr\},
\end{eqnarray}
\begin{eqnarray}\label{ugujjgdhg}
J_{4}(P^{2} + K_{\mu\nu}P^{\mu}P^{\nu}) = \frac{m^{2}}{\mu^{2}}\int_{0}^{1} dx\frac{(1 - x)}{\frac{x(1 - x)(P^{2} + K_{\mu\nu}P^{\mu}P^{\nu})}{\mu^{2}} + \frac{m^{2}}{\mu^{2}}}.
\end{eqnarray}

\begin{acknowledgements}
With great pleasure the authors thank the kind referee for helpful comments. PRSC and MISJ would like to thank Federal University of Piau\'{i} and FAPEAL (Alagoas State Research Foundation), CNPq (Brazilian Funding Agency) for financial support, respectively.
\end{acknowledgements}

\bibliography{apstemplate}

\providecommand{\noopsort}[1]{}\providecommand{\singleletter}[1]{#1}%
\begin{thebibliography}{33}%
\makeatletter
\providecommand \@ifxundefined [1]{%
 \@ifx{#1\undefined}
}%
\providecommand \@ifnum [1]{%
 \ifnum #1\expandafter \@firstoftwo
 \else \expandafter \@secondoftwo
 \fi
}%
\providecommand \@ifx [1]{%
 \ifx #1\expandafter \@firstoftwo
 \else \expandafter \@secondoftwo
 \fi
}%
\providecommand \natexlab [1]{#1}%
\providecommand \enquote  [1]{``#1''}%
\providecommand \bibnamefont  [1]{#1}%
\providecommand \bibfnamefont [1]{#1}%
\providecommand \citenamefont [1]{#1}%
\providecommand \href@noop [0]{\@secondoftwo}%
\providecommand \href [0]{\begingroup \@sanitize@url \@href}%
\providecommand \@href[1]{\@@startlink{#1}\@@href}%
\providecommand \@@href[1]{\endgroup#1\@@endlink}%
\providecommand \@sanitize@url [0]{\catcode `\\12\catcode `\$12\catcode
  `\&12\catcode `\#12\catcode `\^12\catcode `\_12\catcode `\%12\relax}%
\providecommand \@@startlink[1]{}%
\providecommand \@@endlink[0]{}%
\providecommand \url  [0]{\begingroup\@sanitize@url \@url }%
\providecommand \@url [1]{\endgroup\@href {#1}{\urlprefix }}%
\providecommand \urlprefix  [0]{URL }%
\providecommand \Eprint [0]{\href }%
\providecommand \doibase [0]{http://dx.doi.org/}%
\providecommand \selectlanguage [0]{\@gobble}%
\providecommand \bibinfo  [0]{\@secondoftwo}%
\providecommand \bibfield  [0]{\@secondoftwo}%
\providecommand \translation [1]{[#1]}%
\providecommand \BibitemOpen [0]{}%
\providecommand \bibitemStop [0]{}%
\providecommand \bibitemNoStop [0]{.\EOS\space}%
\providecommand \EOS [0]{\spacefactor3000\relax}%
\providecommand \BibitemShut  [1]{\csname bibitem#1\endcsname}%
\let\auto@bib@innerbib\@empty
%</preamble>
\bibitem [{\citenamefont {Kislat}\ and\ \citenamefont
  {Krawczynski}(2015)}]{PhysRevD.92.045016}%
  \BibitemOpen
  \bibfield  {author} {\bibinfo {author} {\bibfnamefont {F.}~\bibnamefont
  {Kislat}}\ and\ \bibinfo {author} {\bibfnamefont {H.}~\bibnamefont
  {Krawczynski}},\ }\href@noop {} {\bibfield  {journal} {\bibinfo  {journal}
  {Phys. Rev. D}\ }\textbf {\bibinfo {volume} {92}},\ \bibinfo {pages} {045016}
  (\bibinfo {year} {2015})}\BibitemShut {NoStop}%
\bibitem [{\citenamefont {Amelino-Camelia}\ \emph {et~al.}(2015)\citenamefont
  {Amelino-Camelia}, \citenamefont {Guetta},\ and\ \citenamefont
  {Piran}}]{0004-637X-806-2-269}%
  \BibitemOpen
  \bibfield  {author} {\bibinfo {author} {\bibfnamefont {G.}~\bibnamefont
  {Amelino-Camelia}}, \bibinfo {author} {\bibfnamefont {D.}~\bibnamefont
  {Guetta}}, \ and\ \bibinfo {author} {\bibfnamefont {T.}~\bibnamefont
  {Piran}},\ }\href@noop {} {\bibfield  {journal} {\bibinfo  {journal} {ApJ}\
  }\textbf {\bibinfo {volume} {806}},\ \bibinfo {pages} {269} (\bibinfo {year}
  {2015})}\BibitemShut {NoStop}%
\bibitem [{\citenamefont {Ribeiro}\ \emph {et~al.}(2015)\citenamefont
  {Ribeiro}, \citenamefont {Passos}, \citenamefont {Furtado},\ and\
  \citenamefont {Nascimento}}]{doi:10.1142/S0217751X15500724}%
  \BibitemOpen
  \bibfield  {author} {\bibinfo {author} {\bibfnamefont {L.~R.}\ \bibnamefont
  {Ribeiro}}, \bibinfo {author} {\bibfnamefont {E.}~\bibnamefont {Passos}},
  \bibinfo {author} {\bibfnamefont {C.}~\bibnamefont {Furtado}}, \ and\
  \bibinfo {author} {\bibfnamefont {J.~R.}\ \bibnamefont {Nascimento}},\
  }\href@noop {} {\bibfield  {journal} {\bibinfo  {journal} {Int. J. Mod. Phys.
  A}\ }\textbf {\bibinfo {volume} {30}},\ \bibinfo {pages} {1550072} (\bibinfo
  {year} {2015})}\BibitemShut {NoStop}%
\bibitem [{\citenamefont {van Tilburg}\ and\ \citenamefont {van
  Veghel}(2015)}]{vanTilburg2015236}%
  \BibitemOpen
  \bibfield  {author} {\bibinfo {author} {\bibfnamefont {J.}~\bibnamefont {van
  Tilburg}}\ and\ \bibinfo {author} {\bibfnamefont {M.}~\bibnamefont {van
  Veghel}},\ }\href@noop {} {\bibfield  {journal} {\bibinfo  {journal} {Phys.
  Lett. B}\ }\textbf {\bibinfo {volume} {742}},\ \bibinfo {pages} {236}
  (\bibinfo {year} {2015})}\BibitemShut {NoStop}%
\bibitem [{\citenamefont {Anacleto}\ \emph {et~al.}(2012)\citenamefont
  {Anacleto}, \citenamefont {Brito},\ and\ \citenamefont
  {Passos}}]{PhysRevD.86.125015}%
  \BibitemOpen
  \bibfield  {author} {\bibinfo {author} {\bibfnamefont {M.~A.}\ \bibnamefont
  {Anacleto}}, \bibinfo {author} {\bibfnamefont {F.~A.}\ \bibnamefont {Brito}},
  \ and\ \bibinfo {author} {\bibfnamefont {E.}~\bibnamefont {Passos}},\
  }\href@noop {} {\bibfield  {journal} {\bibinfo  {journal} {Phys. Rev. D}\
  }\textbf {\bibinfo {volume} {86}},\ \bibinfo {pages} {125015} (\bibinfo
  {year} {2012})}\BibitemShut {NoStop}%
\bibitem [{\citenamefont {Ferrero}\ and\ \citenamefont
  {Altschul}(2011)}]{PhysRevD.84.065030}%
  \BibitemOpen
  \bibfield  {author} {\bibinfo {author} {\bibfnamefont {A.}~\bibnamefont
  {Ferrero}}\ and\ \bibinfo {author} {\bibfnamefont {B.}~\bibnamefont
  {Altschul}},\ }\href@noop {} {\bibfield  {journal} {\bibinfo  {journal}
  {Phys. Rev. D}\ }\textbf {\bibinfo {volume} {84}},\ \bibinfo {pages} {065030}
  (\bibinfo {year} {2011})}\BibitemShut {NoStop}%
\bibitem [{\citenamefont {Carvalho}(2013)}]{Carvalho2013850}%
  \BibitemOpen
  \bibfield  {author} {\bibinfo {author} {\bibfnamefont {P.~R.~S.}\
  \bibnamefont {Carvalho}},\ }\href@noop {} {\bibfield  {journal} {\bibinfo
  {journal} {Phys. Lett. B}\ }\textbf {\bibinfo {volume} {726}},\ \bibinfo
  {pages} {850} (\bibinfo {year} {2013})}\BibitemShut {NoStop}%
\bibitem [{\citenamefont {Carvalho}(2014)}]{Carvalho2014320}%
  \BibitemOpen
  \bibfield  {author} {\bibinfo {author} {\bibfnamefont {P.~R.~S.}\
  \bibnamefont {Carvalho}},\ }\href@noop {} {\bibfield  {journal} {\bibinfo
  {journal} {Phys. Lett. B}\ }\textbf {\bibinfo {volume} {730}},\ \bibinfo
  {pages} {320} (\bibinfo {year} {2014})}\BibitemShut {NoStop}%
\bibitem [{\citenamefont {Vieira}\ and\ \citenamefont
  {Carvalho}(2014)}]{EurophysLett10821001}%
  \BibitemOpen
  \bibfield  {author} {\bibinfo {author} {\bibfnamefont {W.~C.}\ \bibnamefont
  {Vieira}}\ and\ \bibinfo {author} {\bibfnamefont {P.~R.~S.}\ \bibnamefont
  {Carvalho}},\ }\href@noop {} {\bibfield  {journal} {\bibinfo  {journal}
  {Europhys. Lett.}\ }\textbf {\bibinfo {volume} {108}},\ \bibinfo {pages}
  {21001} (\bibinfo {year} {2014})}\BibitemShut {NoStop}%
\bibitem [{\citenamefont {Carvalho}(2016)}]{doi:10.1142/S0217979215502598}%
  \BibitemOpen
  \bibfield  {author} {\bibinfo {author} {\bibfnamefont {P.~R.~S.}\
  \bibnamefont {Carvalho}},\ }\href@noop {} {\bibfield  {journal} {\bibinfo
  {journal} {Int. J. Mod. Phys. B}\ }\textbf {\bibinfo {volume} {30}},\
  \bibinfo {pages} {1550259} (\bibinfo {year} {2016})}\BibitemShut {NoStop}%
\bibitem [{\citenamefont {Vieira}\ and\ \citenamefont
  {de~Carvalho}(2016)}]{doi:10.1142/S0219887816500493}%
  \BibitemOpen
  \bibfield  {author} {\bibinfo {author} {\bibfnamefont {W.~d.~C.}\
  \bibnamefont {Vieira}}\ and\ \bibinfo {author} {\bibfnamefont {P.~R.~S.}\
  \bibnamefont {de~Carvalho}},\ }\href@noop {} {\bibfield  {journal} {\bibinfo
  {journal} {Int. J. Geom. Methods Mod. Phys.}\ }\textbf {\bibinfo {volume}
  {13}},\ \bibinfo {pages} {1650049} (\bibinfo {year} {2016})}\BibitemShut
  {NoStop}%
\bibitem [{\citenamefont {Pelissetto}\ and\ \citenamefont
  {Vicari}(2002)}]{Pelissetto2002549}%
  \BibitemOpen
  \bibfield  {author} {\bibinfo {author} {\bibfnamefont {A.}~\bibnamefont
  {Pelissetto}}\ and\ \bibinfo {author} {\bibfnamefont {E.}~\bibnamefont
  {Vicari}},\ }\href@noop {} {\bibfield  {journal} {\bibinfo  {journal} {Phys.
  Rep.}\ }\textbf {\bibinfo {volume} {368}},\ \bibinfo {pages} {549} (\bibinfo
  {year} {2002})}\BibitemShut {NoStop}%
\bibitem [{\citenamefont {Moukouri}\ and\ \citenamefont
  {Eidelstein}(2012)}]{PhysRevB.86.155112}%
  \BibitemOpen
  \bibfield  {author} {\bibinfo {author} {\bibfnamefont {S.}~\bibnamefont
  {Moukouri}}\ and\ \bibinfo {author} {\bibfnamefont {E.}~\bibnamefont
  {Eidelstein}},\ }\href@noop {} {\bibfield  {journal} {\bibinfo  {journal}
  {Phys. Rev. B}\ }\textbf {\bibinfo {volume} {86}},\ \bibinfo {pages} {155112}
  (\bibinfo {year} {2012})}\BibitemShut {NoStop}%
\bibitem [{\citenamefont {Nishiyama}(2005)}]{PhysRevE.71.046112}%
  \BibitemOpen
  \bibfield  {author} {\bibinfo {author} {\bibfnamefont {Y.}~\bibnamefont
  {Nishiyama}},\ }\href@noop {} {\bibfield  {journal} {\bibinfo  {journal}
  {Phys. Rev. E}\ }\textbf {\bibinfo {volume} {71}},\ \bibinfo {pages} {046112}
  (\bibinfo {year} {2005})}\BibitemShut {NoStop}%
\bibitem [{\citenamefont {Codello}\ and\ \citenamefont
  {D'Odorico}(2013)}]{PhysRevLett.110.141601}%
  \BibitemOpen
  \bibfield  {author} {\bibinfo {author} {\bibfnamefont {A.}~\bibnamefont
  {Codello}}\ and\ \bibinfo {author} {\bibfnamefont {G.}~\bibnamefont
  {D'Odorico}},\ }\href@noop {} {\bibfield  {journal} {\bibinfo  {journal}
  {Phys. Rev. Lett.}\ }\textbf {\bibinfo {volume} {110}},\ \bibinfo {pages}
  {141601} (\bibinfo {year} {2013})}\BibitemShut {NoStop}%
\bibitem [{\citenamefont {Butti}\ and\ \citenamefont
  {Toldin}(2005)}]{Butti2005527}%
  \BibitemOpen
  \bibfield  {author} {\bibinfo {author} {\bibfnamefont {A.}~\bibnamefont
  {Butti}}\ and\ \bibinfo {author} {\bibfnamefont {F.~P.}\ \bibnamefont
  {Toldin}},\ }\href@noop {} {\bibfield  {journal} {\bibinfo  {journal} {Nuc.
  Phys. B}\ }\textbf {\bibinfo {volume} {704}},\ \bibinfo {pages} {527}
  (\bibinfo {year} {2005})}\BibitemShut {NoStop}%
\bibitem [{\citenamefont {Patrascioiu}\ and\ \citenamefont
  {Seiler}(1996)}]{PhysRevB.54.7177}%
  \BibitemOpen
  \bibfield  {author} {\bibinfo {author} {\bibfnamefont {A.}~\bibnamefont
  {Patrascioiu}}\ and\ \bibinfo {author} {\bibfnamefont {E.}~\bibnamefont
  {Seiler}},\ }\href@noop {} {\bibfield  {journal} {\bibinfo  {journal} {Phys.
  Rev. B}\ }\textbf {\bibinfo {volume} {54}},\ \bibinfo {pages} {7177}
  (\bibinfo {year} {1996})}\BibitemShut {NoStop}%
\bibitem [{\citenamefont {Oppermann}\ and\ \citenamefont
  {Schmidt}(2008)}]{PhysRevE.78.061124}%
  \BibitemOpen
  \bibfield  {author} {\bibinfo {author} {\bibfnamefont {R.}~\bibnamefont
  {Oppermann}}\ and\ \bibinfo {author} {\bibfnamefont {M.~J.}\ \bibnamefont
  {Schmidt}},\ }\href@noop {} {\bibfield  {journal} {\bibinfo  {journal} {Phys.
  Rev. E}\ }\textbf {\bibinfo {volume} {78}},\ \bibinfo {pages} {061124}
  (\bibinfo {year} {2008})}\BibitemShut {NoStop}%
\bibitem [{\citenamefont {Trugenberger}(2005)}]{Trugenberger2005509}%
  \BibitemOpen
  \bibfield  {author} {\bibinfo {author} {\bibfnamefont {C.~A.}\ \bibnamefont
  {Trugenberger}},\ }\href@noop {} {\bibfield  {journal} {\bibinfo  {journal}
  {Nuc. Phys. B}\ }\textbf {\bibinfo {volume} {716}},\ \bibinfo {pages} {509}
  (\bibinfo {year} {2005})}\BibitemShut {NoStop}%
\bibitem [{\citenamefont {Amit}\ and\ \citenamefont
  {Mart\'in-Mayor}(2005)}]{Amit}%
  \BibitemOpen
  \bibfield  {author} {\bibinfo {author} {\bibfnamefont {D.~J.}\ \bibnamefont
  {Amit}}\ and\ \bibinfo {author} {\bibfnamefont {V.}~\bibnamefont
  {Mart\'in-Mayor}},\ }\href@noop {} {\emph {\bibinfo {title} {Field Theory,
  The Renormalization Group and Critical Phenomena}}}\ (\bibinfo  {publisher}
  {World Scientific Pub Co Inc},\ \bibinfo {year} {2005})\BibitemShut {NoStop}%
\bibitem [{\citenamefont {Carvalho}\ and\ \citenamefont
  {Leite}(2013)}]{:/content/aip/journal/jmp/54/9/10.1063/1.4819259}%
  \BibitemOpen
  \bibfield  {author} {\bibinfo {author} {\bibfnamefont {P.~R.~S.}\
  \bibnamefont {Carvalho}}\ and\ \bibinfo {author} {\bibfnamefont {M.~M.}\
  \bibnamefont {Leite}},\ }\href@noop {} {\bibfield  {journal} {\bibinfo
  {journal} {J. Math. Phys.}\ }\textbf {\bibinfo {volume} {54}},\ \bibinfo
  {eid} {093301} (\bibinfo {year} {2013})}\BibitemShut {NoStop}%
\bibitem [{\citenamefont {Weinberg}(1960)}]{PhysRev.118.838}%
  \BibitemOpen
  \bibfield  {author} {\bibinfo {author} {\bibfnamefont {S.}~\bibnamefont
  {Weinberg}},\ }\href@noop {} {\bibfield  {journal} {\bibinfo  {journal}
  {Phys. Rev.}\ }\textbf {\bibinfo {volume} {118}},\ \bibinfo {pages} {838}
  (\bibinfo {year} {1960})}\BibitemShut {NoStop}%
\bibitem [{\citenamefont {Wilson}\ and\ \citenamefont
  {Kogut}(1974)}]{Wilson197475}%
  \BibitemOpen
  \bibfield  {author} {\bibinfo {author} {\bibfnamefont {K.~G.}\ \bibnamefont
  {Wilson}}\ and\ \bibinfo {author} {\bibfnamefont {J.}~\bibnamefont {Kogut}},\
  }\href@noop {} {\bibfield  {journal} {\bibinfo  {journal} {Phys. Rep.}\
  }\textbf {\bibinfo {volume} {12}},\ \bibinfo {pages} {75} (\bibinfo {year}
  {1974})}\BibitemShut {NoStop}%
\bibitem [{\citenamefont {Wilson}\ and\ \citenamefont
  {Fisher}(1972)}]{PhysRevLett.28.240}%
  \BibitemOpen
  \bibfield  {author} {\bibinfo {author} {\bibfnamefont {K.~G.}\ \bibnamefont
  {Wilson}}\ and\ \bibinfo {author} {\bibfnamefont {M.~E.}\ \bibnamefont
  {Fisher}},\ }\href@noop {} {\bibfield  {journal} {\bibinfo  {journal} {Phys.
  Rev. Lett.}\ }\textbf {\bibinfo {volume} {28}},\ \bibinfo {pages} {240}
  (\bibinfo {year} {1972})}\BibitemShut {NoStop}%
\bibitem [{\citenamefont {Wilson}(1972)}]{PhysRevLett.28.548}%
  \BibitemOpen
  \bibfield  {author} {\bibinfo {author} {\bibfnamefont {K.~G.}\ \bibnamefont
  {Wilson}},\ }\href@noop {} {\bibfield  {journal} {\bibinfo  {journal} {Phys.
  Rev. Lett.}\ }\textbf {\bibinfo {volume} {28}},\ \bibinfo {pages} {548}
  (\bibinfo {year} {1972})}\BibitemShut {NoStop}%
\bibitem [{\citenamefont {Bogoliubov}\ and\ \citenamefont
  {Parasyuk}(1957)}]{BogoliubovParasyuk}%
  \BibitemOpen
  \bibfield  {author} {\bibinfo {author} {\bibfnamefont {N.~N.}\ \bibnamefont
  {Bogoliubov}}\ and\ \bibinfo {author} {\bibfnamefont {O.~S.}\ \bibnamefont
  {Parasyuk}},\ }\href@noop {} {\bibfield  {journal} {\bibinfo  {journal} {Acta
  Math.}\ }\textbf {\bibinfo {volume} {97}},\ \bibinfo {pages} {227} (\bibinfo
  {year} {1957})}\BibitemShut {NoStop}%
\bibitem [{\citenamefont {Hepp}(1966)}]{Hepp}%
  \BibitemOpen
  \bibfield  {author} {\bibinfo {author} {\bibfnamefont {K.}~\bibnamefont
  {Hepp}},\ }\href@noop {} {\bibfield  {journal} {\bibinfo  {journal} {Commun.
  Math. Phys.}\ }\textbf {\bibinfo {volume} {2}},\ \bibinfo {pages} {301}
  (\bibinfo {year} {1966})}\BibitemShut {NoStop}%
\bibitem [{\citenamefont {Zimmermann}(1969)}]{Zimmermann}%
  \BibitemOpen
  \bibfield  {author} {\bibinfo {author} {\bibfnamefont {W.}~\bibnamefont
  {Zimmermann}},\ }\href@noop {} {\bibfield  {journal} {\bibinfo  {journal}
  {Commun. Math. Phys.}\ }\textbf {\bibinfo {volume} {15}},\ \bibinfo {pages}
  {208} (\bibinfo {year} {1969})}\BibitemShut {NoStop}%
\bibitem [{\citenamefont {Leite}(2003)}]{Phys.Rev.B.67.104415}%
  \BibitemOpen
  \bibfield  {author} {\bibinfo {author} {\bibfnamefont {M.~M.}\ \bibnamefont
  {Leite}},\ }\href@noop {} {\bibfield  {journal} {\bibinfo  {journal} {Phys.
  Rev. B}\ }\textbf {\bibinfo {volume} {67}},\ \bibinfo {pages} {104415}
  (\bibinfo {year} {2003})}\BibitemShut {NoStop}%
\bibitem [{\citenamefont {Leite}(2005)}]{Phys.Rev.B.72.224432}%
  \BibitemOpen
  \bibfield  {author} {\bibinfo {author} {\bibfnamefont {M.~M.}\ \bibnamefont
  {Leite}},\ }\href@noop {} {\bibfield  {journal} {\bibinfo  {journal} {Phys.
  Rev. B}\ }\textbf {\bibinfo {volume} {72}},\ \bibinfo {pages} {224432}
  (\bibinfo {year} {2005})}\BibitemShut {NoStop}%
\bibitem [{\citenamefont {Carvalho}\ and\ \citenamefont
  {Leite}(2009)}]{Carvalho2009178}%
  \BibitemOpen
  \bibfield  {author} {\bibinfo {author} {\bibfnamefont {P.~R.~S.}\
  \bibnamefont {Carvalho}}\ and\ \bibinfo {author} {\bibfnamefont {M.~M.}\
  \bibnamefont {Leite}},\ }\href@noop {} {\bibfield  {journal} {\bibinfo
  {journal} {Ann. Phys.}\ }\textbf {\bibinfo {volume} {324}},\ \bibinfo {pages}
  {178} (\bibinfo {year} {2009})}\BibitemShut {NoStop}%
\bibitem [{\citenamefont {Carvalho}\ and\ \citenamefont
  {Leite}(2010)}]{Carvalho2010151}%
  \BibitemOpen
  \bibfield  {author} {\bibinfo {author} {\bibfnamefont {P.~R.~S.}\
  \bibnamefont {Carvalho}}\ and\ \bibinfo {author} {\bibfnamefont {M.~M.}\
  \bibnamefont {Leite}},\ }\href@noop {} {\bibfield  {journal} {\bibinfo
  {journal} {Ann. Phys.}\ }\textbf {\bibinfo {volume} {325}},\ \bibinfo {pages}
  {151} (\bibinfo {year} {2010})}\BibitemShut {NoStop}%
\bibitem [{\citenamefont {Brezin}\ \emph {et~al.}(1973)\citenamefont {Brezin},
  \citenamefont {Le~Guillou},\ and\ \citenamefont
  {Zinn-Justin}}]{PhysRevD.8.434}%
  \BibitemOpen
  \bibfield  {author} {\bibinfo {author} {\bibfnamefont {E.}~\bibnamefont
  {Brezin}}, \bibinfo {author} {\bibfnamefont {J.~C.}\ \bibnamefont
  {Le~Guillou}}, \ and\ \bibinfo {author} {\bibfnamefont {J.}~\bibnamefont
  {Zinn-Justin}},\ }\href@noop {} {\bibfield  {journal} {\bibinfo  {journal}
  {Phys. Rev. D}\ }\textbf {\bibinfo {volume} {8}},\ \bibinfo {pages} {434}
  (\bibinfo {year} {1973})}\BibitemShut {NoStop}%
\end{thebibliography}%

\end{document}